\documentclass[prb,twocolumn,a4paper,superscriptaddress,showpacs,showkeys]{revtex4}

\usepackage{amssymb}
\usepackage{graphicx,psfrag,color}
\usepackage{times,bm}

\def\be{\begin{equation}}          \def\ee{\end{equation}}
\def\bea{\jot2ex\begin{eqnarray}}         \def\eea{\end{eqnarray}}
\newsavebox{\weg} 
\def\NL{\nonumber \\*}

\def\oml{\langle \omega |} \def\omr{| \omega \rangle }   
\def\dtau{\partial^{}_{\tau}}
\def\vr{{\bf r}} \def\vrs{{\bf r}'}
\def\vx{{\bf x}}  \def\vq{{\bf q}}  \def\vk{{\bf k}} 
\def\up{\uparrow}  \def\dwn{\downarrow}
\def\vn{{\bf n}}
\def\vd{\bm{\delta}}  \def\vA{{\bf A}}
\def\ho#1{(2S - \eta_{#1}^* \eta_{#1}^{})}
\def\halb{\frac{1}{2}}
\def\Sp#1{\halb \; {\rm Tr}\left\{#1\right\}}
\def\Sm{S \! - \! \halb}
\def\eins#1{\openone_{\rule[.5ex]{0ex}{1ex}{#1}}}
\def\re#1{(\ref{#1})}
\def\et{\tilde{\eta}}

\def\eps{\bm{\epsilon}}   \def\epst{\tilde{\bm{\epsilon}}}
\def\sumk#1#2{\frac{1}{{\cal N}^#1} \sum_{#2} }
\def\omk#1{\omega_{#1}}
\def\oq{\omk{\vk}}  \def\oqs{\omk{\vk'}}  
\def\gam#1{\gamma_{#1}}
\def\gq{\gam{\vk}}  \def\gqs{\gam{\vk'}}  \def\gqp{\gam{\vk + \vk'}}
\def\Lnls{{\cal L}_{\mbox{\footnotesize nl$\sigma$m}}}
\def\Loneh{{\cal L}_{\mbox{\footnotesize 1--hole}}}
\def\Goneh{G^{\mbox{\footnotesize 1--hole}}}

\begin{document}

\title{Single--hole dynamics in the $t$-$J$ model}

\author{W. Apel}
\affiliation{Physikalisch-Technische Bundesanstalt,
Bundesallee 100, 38116 Braunschweig, Germany.}

\author{H.-U. Everts}
\author{U. K\"orner}
\affiliation{Institut f\"ur Theoretische Physik, Universit\"at Hannover,
Appelstr.\ 2, 30167 Hannover, Germany.}

\date{\today}

\begin{abstract}
The quasi--particle weight of a single hole in an antiferromagnetic 
background is studied in the semiclassical approximation. 
We start from the $t$-$J$ model, generalize it to arbitrary spin $S$ 
by employing an appropriate coherent state representation for the hole, 
and derive an effective action for the dynamics in the long-wavelength 
low-energy limit.
In the same limit, we find an expression for the single--hole Green's 
function which we evaluate in an $1/S$ expansion. 
Our approach has the advantage of being applicable in one {\em and} 
in two dimensions. 
We find two qualitatively different results in these two cases: 
while in one dimension our results are compatible with a vanishing 
quasi--particle weight, this weight is found to be finite 
in two dimensions, indicating normal quasi-particle behavior 
of the hole in this last case. 
\end{abstract}

\pacs{71.10.Fd, 71.10.Pm, 75.10.Jm}
\keywords{Hole propagation, t-J model}

\maketitle

\section{Introduction}
Although more than 30 years have elapsed since  Brinkman and Rice \cite{BR70} 
published their pioneering work on the propagation of a hole in a quantum 
antiferromagnet, this subject still attracts a lot of attention. 
Central to this problem is the question whether the motion of a hole in a 
strongly correlated system is coherent or not, i.e., whether the hole can 
be considered as a quasi--particle in the sense of Landau's Fermi--liquid 
theory, or whether, as was first suggested by Anderson \cite{A87,A90}, 
it is more appropriate to view the hole as a constituent particle of a 
Luttinger--like liquid. 
For a long time, the interest in this question has been fueled by 
the expectation that an accurate description of the hole dynamics 
in a strongly correlated electron system would be the crucial first step 
towards an understanding of the physics of the cuprate superconductors. 
More recently, this interest has been revived, since the dynamics of 
holes in one-- (1D) and two--dimensional (2D) antiferromagnets has become 
experimentally accessible in angle--resolved photoemission spectroscopy 
(ARPES) \cite{WSMKKGB95,KMSMEUTM96}. 
The key quantity that can in principle be obtained from the results of 
ARPES is the spectral weight $Z$ at the ground state energy $E_G$ of 
the systems.  
A nonvanishing spectral weight, $Z \neq 0$, implies that the hole is 
a quasi--particle, e.g., a spin-polaron consisting of the bare hole 
with a spin polarisation cloud around it. 
On the contrary, a vanishing $Z$ signals that the hole causes a global 
rearrangement of the original ground state such that the exact single--hole 
wave function becomes orthogonal to the wave function of a bare hole 
in the antiferromagnetic ground state. 
ARPES data for the one--dimensional copper oxide chain compound SrCuO${}_2$, 
see Ref.~\onlinecite{KMSMEUTM96},  
and for the two-dimensional layered compound Sr${}_2$CuO${}_2$Cl${}_2$, 
see Ref.~\onlinecite{WSMKKGB95}, show that the hole dynamics differ 
significantly between the 1D and the 2D case (see the discussion in 
Ref.~\onlinecite{KMSMEUTM96}). 
The features of the spectra of the 1D compound fit rather well with 
the concept of spin--charge separation that has emerged from Bethe--Ansatz 
and bosonization studies of the 1D Hubbard model \cite{SP92,PS92,ET02}
and from the exact solution of the $t$-$J$ model at the supersymmetric 
point \cite{SP96,SP98}.   
Theoretically, spin--charge separation implies that for long wavelengths, 
the Hamiltonian of a hole can be decomposed into two commuting parts, 
$H=H_h+H_s$, where both the holon part $H_h$ and the spinon part $H_s$, 
are free-fermion Hamiltonians \cite{SP96,SN97,SP98}. 
As a consequence, the single--hole Green's function takes the form 
\be
 G(p,\tau) \;=\; \int\!\frac{dQ}{2\pi} \;\; G_h (p-Q,\tau) \; Z(Q,\tau) \;.
\ee

Here, $G_h(k,\tau)$ is the propagator of a free holon, whereas the 
function $Z(Q,\tau)$ is entirely determined by the spinon dynamics; 
$Q$ is the spinon momentum. 
While the holon Green's function $G_h$ is a free-fermion propagator indeed, 
$Im \; G_h(k,\omega)=\pi \delta (\omega - \varepsilon_h(k))$ with 
$\varepsilon_h(k)$ the holon kinetic energy, the spinon contribution 
$Z(Q,\tau)$ is a highly nontrivial singular function.
Technically, this last feature has its origin in the fact that the original 
hole operator $c_{i\sigma}$ does not simply factorize into a holon and 
a spinon operator, $c_{i\sigma}\neq h^{\dagger}_i s_i$. 
Rather, in the representation of $c_{i\sigma}$ by $h^{\dagger}_i$ and  
$s_i$, an additional phase factor $\exp(i \theta_s)$ is needed, 
where the phase $\theta_s$ depends on the spinon operators $s_j$ only. 
This phase factor accounts for the {\em phase--string} 
effect \cite{SCW96,WSCT97,SN97}: 
as the physical hole moves through the (quasi)--ordered spin background, 
it trails a string of overturned spins behind it. 
Using the {\em phase--string} picture, Suzuura and Nagaosa \cite{SN97} 
have obtained a qualitatively correct result for the single--hole 
spectral function in 1D. 
We note, however, that Sorella and Parola \cite{SP96,SP98} have derived 
their detailed results for the spinon function $Z(Q,\tau)$ in 1D 
without making use of a {\em phase--string} picture. 
Thus in 1D, the different approaches lead to the same conclusion: 
there is spin--charge separation in this case, and the hole Green's    
function shows Luttinger--liquid--like behavior. 
A recent quantum Monte--Carlo (QMC) simulation of the single--hole 
dynamics in the one--dimensional $t$-$J$ model has confirmed these 
results \cite{BAM00a}. 

In two dimensions, the situation is less clear. 
Many studies of the hole dynamics in the two--dimensional $t$-$J$ model 
have been based on an effective Hamiltonian which describes the hole as 
a spinless fermion that is coupled linearly to the magnon excitations 
of the antiferromagnetic background. 
In obtaining the hole Green's function, the coupling between the hole 
and the magnons is then treated in the self--consistent Born 
approximation (SCBA), and a normal quasi--particle peak is found in 
the single-hole spectral function \cite{SVR88,KLR89,MRSV91,LM91,MH91}. 
It must be emphasized that the SCBA is a perturbative method so that 
the quasi--particle picture is inherent in this method. 
However, a finite quasi-particle weight of  a hole has also been found 
in a non-perturbative large-spin study of the 2D t-J model \cite{ASP95}.
Moreover, a recent QMC simulation \cite{BAM00b} lends support to these 
results in a wide range of the model parameter $J/t$ 
and of the momentum $k$ of the propagating hole. 
But, remarkably, two theoretical predictions  by Sorella \cite{S96} 
concerning the behavior of the quasi--particle weight Z(Q) for $J/t=2$ 
at the antiferromagnetic wave vector ${\bf Q} = (\pi,\pi)$ are 
less well confirmed by the simulations in Ref.~\onlinecite{BAM00b}. 
Apart from this seemingly minor discrepancy between QMC simulation 
and certain theoretical predictions, there is further work which casts 
doubt on the validity of the quasi--particle picture for a hole 
in the two--dimensional $t$-$J$ model: 
results from a high temperature series expansion for the momentum 
distribution of the particles in the $t$-$J$ model show a violation of 
Luttinger's theorem \cite{PLS98}. 
This means that the ground state of the 2D $t$-$J$ model is not 
connected adiabatically to the ground state of the noninteracting model 
as would be the case for a normal Fermi--liquid consisting of 
quasi--particles. 
More directly, it is claimed by Weng {\em et al.} 
\cite{SCW96,WSCT97,WMST01} that the phase--string picture, which explains 
the vanishing of the quasi--particle weight in 1D,  
applies in 2D too and has the same effect there.
Despite  this apparent failure of the quasi--particle picture, 
these authors suggest in Ref.~\onlinecite{WMST01} that the spectral 
features seen in ARPES can be explained satisfactorily in the framework 
of the $t$-$J$ model by their theory. 
However, in a very recent QMC study Mishchenko {\em et~al.} \cite{MPS01} 
contradict this suggestion, as they confirm the quasi--particle picture 
already found by Brunner {\em et~al.} \cite{BAM00b}.

In view of these contradictions between numerical and analytic analyses, 
we propose a new analytic approach to the calculation of the 
quasi--particle weight $Z$ of a single hole in the $t$-$J$ model. 
Our method has the advantage of being applicable in arbitrary space 
dimensions and is thus suitable to detect differences between 
the 1D and the 2D case.   
In 1D, our results are compatible with a power--law decay of 
the weight $Z$ as the system size $L$ increases to infinity, 
$Z\sim L^{-2X}$. 
On the contrary in 2D, this weight remains finite in the thermodynamic 
limit indicating that the quasi--particle picture is valid in this case.

The paper is organised as follows: 
Section II contains the derivation of the effective action of 
a single--hole in an antiferromagnetic background with the method of 
coherent states \cite{W88}. 
In the later parts of the paper, we wish to employ a semiclassical 
expansion, i.e., an expansion in powers of the inverse spin length $1/S$. 
For this reason, we generalise the conventional $t$-$J$ model in which 
every lattice site is either occupied by a spin $\frac{1}{2}$
particle or empty, to the case where the lattice sites are occupied 
either by an object with arbitrary but fixed spin $S$ (particle) 
or by an object with spin $S-\frac{1}{2}$ (hole) \cite{AL91}. 
We implement this modification by an appropriate choice of the 
coherent--state representation for the particles which at the same time 
satisfies explicitely the constraint of forbidden double occupancy.
In describing the local coupling of the hole to the spin degrees of 
freedom, it turns out to be crucial to use local $SU(2)$ fields 
as the basic variables of the effective action instead of working 
with the local sublattice magnetization. 
In Section III, we employ the effective action of Section II to 
obtain a path--integral representation for the single--hole Green's 
function. 
Our aim is to extract from this function an expression for the 
quasi--particle weight $Z$ of the hole which exhibits explicitly 
the dependence of $Z$ on the system parameters, in particular its 
dependence on the linear system size $L$. 
To achieve this goal, we have to take recourse to the semiclassical 
expansion. 
In Section IV, we present and discuss our final results. 
Technicalities of the developments in Sections II and III are deferred 
to Appendices A and B.

\section{Path--integral representation}

In this Section, we derive an effective action for the motion 
of a single hole in an antiferromagnet.  
The underlying Hamiltonian is that of the $t$-$J$ model, 

\bea
{\cal H}&=& -t \sum_{\textstyle 
 \stackrel {<\vr,\vrs>}{\scriptstyle \rule{0pt}{1ex}\sigma}}
\hat{P} \; \left( c_{\vr,\sigma}^{\dagger} c_{\vrs \!,\sigma}^{} 
        + {\it h.c.} \right) \; \hat{P} \NL
&& + J \sum_{<\vr,\vrs>} \hat{\bf S}_{\vr}\cdot \hat{\bf S}_{\vrs} \;\;.
\label{t-J}
\eea
Here $c_{\vr,\sigma}^{\dagger}$ and $ c_{\vr \!,\sigma}$ are the creation 
and annihilation operators for fermion states at site ${\vr}$ with 
spin projection $\sigma$, and
$\hat{\bf S}_{\vr} = \frac{1}{2}c_{\vr,\alpha}^{\dagger}\,
{\bm \sigma}_{\alpha,\beta}^{}\,c_{\vr,\beta}^{}$ is the spin of the fermion. 
$\hat{P}$ projects onto states where each lattice site is either empty or 
singly occupied.
$<\!\vr,\vrs\!>$ denotes nearest neighbor sites on a hypercubic lattice with 
lattice constant $a$ in $D$ dimensions. 
The main difficulty in analysing the properties of \re{t-J} lies in the 
strong interactions induced by the exclusion of doubly occupied sites.
We deal with this difficulty by using an appropriate set of coherent states 
(cf.~Refs.~\onlinecite{ZFG90}, \onlinecite{W88}) which takes these 
constraints into account explicitely. 
This leads to the following path-integral representation of the partition 
function ${\cal Z}$:

\be
{\cal Z} \;=\; \int \!{\cal D}[\omega] \; e^{-\int_0^{\beta}d\tau \;
 \left\{ \oml \dtau \omr \;+\; \oml{\cal H} \omr \right\} }  \;\; .
\label{zet}
\ee

The coherent states $\omr$ are introduced and discussed in the 
Appendix \ref{Koh-Zust}. 
They are parameterized by two fields; the anticommuting fields 
$\eta_{\vr}^{}$, $\eta_{\vr}^*$ which describe the hole, 
and the commuting field  $g_{\vr} \in SU(2)$ which determines the orientation 
of the spin (see Eq.~\re{nvec} below).
As is detailed in the Appendix, our coherent states are constructed such 
that they allow us to generalise to arbitrary spin $S$: 
Each lattice site is occupied either by an object with spin $S$ (particle) 
or an object with spin $\Sm$ (hole) \cite{AL91}. 
Then, the hopping term $t$ in Eq.~\re{t-J} exchanges spin $S$ objects and  
spin $\Sm$ objects on neighbouring lattice sites. 
The corresponding matrix elements are worked out in the Appendix \ref{Koh-Zust}. 
In the parametrisation by $\eta$ and $g$, the term entering the kinetic part 
of the action in Eq.~\re{zet} takes the form

\be
 \oml \dtau \omr \;=\; \sum_{\vr} \left\{ \ho{\vr} 
 \left( g_{\vr}^{\dag} \dtau g_{\vr}^{} \right)_{\up \up}
 \;+\; \eta_{\vr}^* \dtau \eta_{\vr}^{} \right\} \;,   \label{Skin}
\ee

while in the Hamiltonian part 

{\arraycolsep0ex
\bea
 \oml {\cal H} \omr \;=&& \sum_{<\vr,\vrs>} \left\{ 2S\; t
 \left[ \eta_{\vr}^* \eta_{\vrs}^{} 
  \left( g_{\vrs}^{\dag} g_{\vr}^{}\right)_{\up \up} 
  \;+\; (\vr \leftrightarrow \vrs ) \right]  \right. \NL
  && \left. + \frac{J}{4} \ho{\vr} \ho{\vrs} \;
  \vn_{\vr} \cdot \vn_{\vrs} \right\}   \, .       \label{Sham}
\eea}

In the case $S=\halb$, everything reduces to the ordinary $t$-$J$ model.
The unit vector

\be
 \vn_{\vr} \;=\; \Sp{ \bm{\sigma} \;\, g_{\vr}^{} \, \sigma_z\,
  g_{\vr}^{\dag} }    \label{nvec}   
\ee

which occurs in the Heisenberg term in Eq.~\ref{Sham} 
points in the direction of the expectation value of the spin. 
The magnitude of the spin is $S$, or $\Sm$, depending on the presence 
of a hole which is expressed by the occupation number 
$\eta_{\vr}^*\eta_{\vr}^{}$.
The amplitude of the hopping term $\eta_{\vr}^* \eta_{\vrs}^{}$ in 
Eq.~\re{Sham} is modulated by the actual state of the background spin, and 
this 
constitutes the main coupling between the hole-- and the spin--subsystems.  
Note that, while the Heisenberg part depends on $g_{\vr}$ only through 
the bilinear term \re{nvec}, the $SU(2)$--fields $g_{\vr}$ at the individual 
lattice sites are needed to express the hopping term (see the related 
discussion of symmetries of the action in Ref.~\onlinecite{AEK98}). 

\begin{figure}
\begin{center}
\fboxsep0.3ex
\psfrag{r1}{\colorbox{white}{$\scriptstyle \vx$}}   
\psfrag{r2}{}   
\psfrag{r3}{\colorbox{white}{$\scriptstyle \vx + \vd$}} 
\psfrag{g1}{\colorbox{white}{$g^{}_{\vx}$}}   
\psfrag{g2}{\colorbox{white}{$g^{\prime}_{\vx}$}}
\includegraphics[width=7cm]{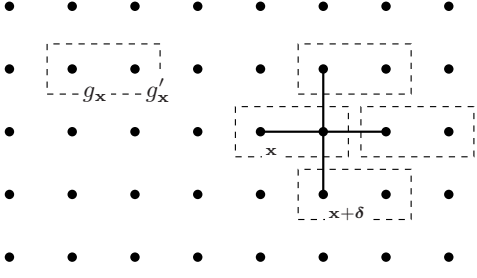}
\end{center}
\caption{\label{gitter} Partitioning of the square lattice into plaquettes}
\end{figure}

Proceeding towards the derivation of an effective action, we now divide the 
lattice into plaquettes each of which  contains a pair of neighbouring sites 
of the two sublattices. 
We label these with $\vx$, the lattice points of the A-sublattice.
Then, the plaquette $\vx$ contains the fields $g^{}_{\vx}$ and 
$g^{\prime}_{\vx} \equiv g_{\vr + a {\hat e}_x}$, on the A-- and B--site, 
respectively, see Fig.~\ref{gitter}.

In the conventional description of a two-sublattice antiferromagnet, 
one would introduce now the magnetisation 
${\bf M} = {\bf S}_A + {\bf S}_B$ and the sublattice magnetisation 
${\bf N} = {\bf S}_A - {\bf S}_B$ of each plaquette as new fields and 
integrate subsequently over $\bf M$.

In our $SU(2)$ description, we introduce the $SU(2)$--field 

\be
 M_{\vx} \;:=\; g^{\prime\dag}_{\vx} \; g^{}_{\vx} \;\;  ;
\ee

we choose the $SU(2)$ fields $g^{}_{\vx}$ and $M_{\vx}$ as our new variables 
and, finally, we will integrate over the fields $M_{\vx}$.
A useful parametrization of $M$ in terms of three real variables $\alpha$, 
$\Re \,u$, and $\Im \,u$ is
 
\be
 M \;=\; 
 \left( \begin{array}{lc} e^{i\alpha} & 0 \\ 0 & e^{-i\alpha} \end{array}\right)
 \; 
 \left( \begin{array}{lc} u & i \\ i & u^* \end{array}\right)
 / \sqrt{1 + |u|^2}
\ee

(here, the label $\vx$ has been suppressed).
Then, for an arbitrary phase $\alpha$, the point $u=0$ corresponds to the 
antiferromagnetic alignment of the spins within the plaquette, since

\bea
 \vn^{}_{\vx} \cdot \vn^{\prime}_{\vx} \;=\;&&
\Sp{ M^{\dag}_{\vx} \sigma_z M^{}_{\vx} \sigma_z } \NL
 \;=\;&& -1 + 2 |u_{\vx}|^2 \;+\; {\cal O}(u_{\vx}^4)  \;. 
\eea

Next, we express all terms in the action by $M_{\vx}$ and $g^{}_{\vx}$; 
we start with the Hamiltonian part. 
The vector between nearest neighbor plaquettes is denoted by
$\vd = 2a{\hat e}_x, a{\hat e}_x \pm a {\hat e}_y, \dots \; (\mbox{or}\; 0)$, 
see Fig.~\ref{gitter}.
Then, the hopping term between two plaquettes in Eq.~\re{Sham} contains the 
expression

\be
g^{\prime\dag}_{\vx} \; g^{}_{\vx + \vd} \;=\;
 M_{\vx} \;\; g^{\dag}_{\vx} \; g^{}_{\vx + \vd}  \;.
\ee

Since we are aiming at a gradient expansion of the  $SU(2)$ fields in the 
Hamiltonian part of the action, we replace 
$g^{\dag}_{\vx} \; g^{}_{\vx + \vd}$ in the hopping term in Eq.~\re{Sham} by

\be
  g^{\dag}_{\vx} \; g^{}_{\vx + \vd} \;=:\; 1 + i \vd \cdot \vA_{\vx}  \; ,
\ee

where $\vA_{\vx}$  becomes the spatial derivative 

\be
  \vA_{\vx} \;=\; -i \; g^{\dag}_{\vx} \;\nabla \; g^{}_{\vx}  \;.
\ee 
in the continuum limit.
The Heisenberg term in Eq.~\re{Sham} can also be conveniently expressed  
by $M_{\vx}$ and $\vA_{\vx}$: 

{\arraycolsep-0.5ex
\bea
 && \vn^{}_{\vx+\vd} \cdot \vn^{\prime}_{\vx} \NL 
 &&=
 \Sp{ g^{\dag}_{\vx + \vd} \; g^{\prime}_{\vx} \;\sigma_z\; 
  g^{\prime\dag}_{\vx} \; g^{}_{\vx + \vd} \;\sigma_z} \NL 
 &&=
 \Sp{\left[ 1 + i \vd \cdot \vA_{\vx} \right]^{\dag} M^{\dag}_{\vx} \;\sigma_z
  \; M^{}_{\vx} \left[ 1 + i \vd \cdot \vA_{\vx} \right] \;\sigma_z } . 
\eea }

In the term entering the kinetic part of the action, Eq.~\re{Skin}, 
we introduce the time derivative 

\be
  A^{\tau}_{\vx} \;:=\; -i \;g^{\dag}_{\vx} \;\dtau\; g^{}_{\vx} \;.
\ee

Then, we have on the B--sites, 

\be
  g^{\prime\dag}_{\vx} \;\dtau\; g^{\prime}_{\vx} \;=\; 
  M^{}_{\vx} \; i A^{\tau}_{\vx} \; M^{\dag}_{\vx} \;+\;
  M^{}_{\vx} \;\dtau\; M^{\dag}_{\vx}  \;.
\ee

Next, we collect all terms and  expand the expressions in the exponent of 
the functional integral for $\cal Z$, Eq.~\re{zet}, up to the second order 
in $u_{\vx}$. 
Finally, we perform the Gaussian integration over $u_{\vx}$ and identify 
the effective action ${\cal A} = \int_0^{\beta}d\tau {\cal L}$ as the exponent 
in the resulting path--integral. 
The remaining integration fields are defined on the A--sublattice: 
on each plaquette there is one  commuting field $g^{}_{\vx} \in SU(2)$ 
for the remaining spin--degrees of freedom, 
and there are two anticommuting fields  $\eta^{}_{\vx}$ and 
$\et_{\vx} = e^{i \,\alpha_{\vx}} \; \eta^{}_{\vr + a {\hat e}_x}$
on each plaquette which describe the holes.
Note that the phase $\alpha_{\vx}$ has been absorbed in the redefinition of 
the hole field on the B--sublattice; it disappears from the effective action.
We work in the leading order in the gradients, $\vA_{\vx}$ and $A^{\tau}_{\vx}$
so that the effective Lagrangian $\cal L$ is accurate in second order 
in these gradients.
$\cal L$ can now be written as a sum of terms, each representing the 
contribution of a fixed number of holes. 
We retain only the zero- and one-hole contributions:
${\cal L}=-2{\cal N} DJS^2 + \Lnls + \Loneh + \cdots$.
The first term is the energy of the classical antiferromagnetic state 
of a system with $\cal N$ plaquettes.
The fluctuations of the spins in the absence of holes are governed in the 
continuum limit by the well--known non--linear $\sigma$--model 
for the unit vector $\vn_{\vx}$

\bea
\Lnls =
 && \int \!\! \frac{d^D x}{2 a^D} \; \Big[
  \frac{1}{4DJ} \dot{\vn}_{\vx} ^2   
 + JS^2 a^2 \sum_{j=1}^D \left( \partial^{}_j \vn_{\vx} \right)^2 \Big] \NL
&& + \, i \, \frac{S}{2} \; \int \!\! \frac{d^D x}{a^{(D-1)}} \;\; 
    \vn_{\vx} \, \cdot \, \dot{\vn}_{\vx} \, \times \, \partial^{}_1 \vn_{\vx}  
   \label{nlsm}
\eea

The first two terms in Eq.~\re{nlsm} are temporal and spatial fluctuations, 
respectively ($\dot{\vn}_{\vx} \equiv \dtau \; \vn_{\vx}$).
The last term is the famous topological term which yields a zero mass for the 
fluctuations for half--integer spins in one dimension; for $D > 1$, this term 
is not effective (see, e.g., the discussion in Ref.~\onlinecite{DR88}). 
We wish to emphasise that we obtain here exactly the same result, 
Eq.~\re{nlsm}, 
as in the conventional approach in which one uses the magnetisation 
${\bf M} = {\bf S}_A + {\bf S}_B$ and the  sublattice 
magnetisation as ${\bf N} = {\bf S}_A - {\bf S}_B$ as plaquette variables and 
then integrates over ${\bf M}$.
However, our approach which uses the $SU(2)$--fields $g_{\vx}$ as basic 
variables turns out to be superior when the steps leading to $\Lnls$ 
have to be performed for the hopping term of Eq.~\re{Sham}. 
From the procedure described above, we get

\bea
\lefteqn{ \Loneh =}&& \NL 
&&\sum_{\vx} \Big[
  \eta^*_{\vx} \left( \dtau - i A^{\tau}_{\vx \, \up\up}  \right) \eta^{}_{\vx}
  \;+\;
  \et^{*}_{\vx} \left( \dtau - i A^{\tau}_{\vx \, \dwn\dwn} \right) \et^{}_{\vx}
   \Big] \NL
&& + \sum_{\vx} \Big[ \eta^*_{\vx} \;h^{aa}_{\vx}\; \eta^{}_{\vx}
  \;+\; \et^*_{\vx} \;h^{bb}_{\vx}\; \et^{}_{\vx} \Big]  \NL
&& + \sum_{\vx,\vd} \Big[ \eta^*_{\vx+\vd} \;h^{ab}_{\vx+\vd,\vx}\; \et^{}_{\vx}
  \;+\; \et^*_{\vx} \;h^{ba}_{\vx,\vx+\vd}\; \eta^{}_{\vx+\vd} \Big]
 \label{l-one-hole}
\eea
where $h^{\alpha\beta}_{\cdots}$ are the matrix elements of the effective 
Hamiltonian for the hole in the background of the $g^{}_{\vx}$--fields. 
The diagonal elements are expressed in terms of the vector $\vn_{\vx}$ 
(bilinear in $g^{\dag}_{\vx}$ and $g^{}_{\vx}$, see Eq.~\re{nvec}) as follows 

\bea
h^{aa}_{\vx} &=& DJS 
   +\frac{1}{8DJS} \dot{\vn}_{\vx} ^2   
   - \frac{J S}{16D} \sum_{\vd , \vd'} (\vn_{\vx+\vd} - \vn_{\vx+\vd'})^2
\NL
h^{bb}_{\vx} &=& DJS 
   -\frac{1}{8DJS} \dot{\vn}_{\vx} ^2 
   - \frac{J S}{16D} \sum_{\vd , \vd'} (\vn_{\vx+\vd} - \vn_{\vx+\vd'})^2 \NL
&& - i \, \frac{1}{2} \; \vn_{\vx} \, \cdot \, \dot{\vn}_{\vx} \, \times \, 
     \frac{1}{2D} \sum_{\vd}(\vn_{\vx+\vd} - \vn_{\vx}) \;. \label{h-diag}
\eea

The off--diagonal elements describe hopping from the A--sublattice to 
the B--sublattice and vice versa. 
They can only be expressed with the aid of the $SU(2)$--field $g^{}_{\vx}$ 
itself:

\bea
h^{ab}_{\vx+\vd,\vx} &=& \frac{-it}{DJ}
      \Big[ g^{\dag}_{\vx} \;\dtau\; g^{}_{\vx} + J S
      \sum_{\vd'} \left( g^{\dag}_{\vx+\vd}-g^{\dag}_{\vx+\vd'} \right)
      g^{}_{\vx}\Big]_{\dwn\up}    \NL  
h^{ba}_{\vx,\vx+\vd} &=& \frac{it}{DJ}
    \Big[ g^{\dag}_{\vx} \;\dtau\; g^{}_{\vx} + J S 
      \sum_{\vd'} g^{\dag}_{\vx} 
      \left( g^{}_{\vx+\vd} - g^{}_{\vx+\vd'} \right) \Big]_{\up\dwn} \;. \NL
\label{h-ausser-diag}
\eea

The preceding equations complete the derivation of the effective action for 
a hole in the background of the spin--fields $g^{}_{\vx}$. 
The approximations made so far were (i) the Gaussian integration over the 
variables $u_{\vx}$ which describe the deviation from the antiferromagnetic 
order within a plaquette and (ii) the gradient expansion in temporal and 
spatial derivatives. 
Thus, the Euler angles defining the fields $g^{}_{\vx}$ should be smooth 
in space and time, but there is no limitation on their variation   
over large distances in space or time.

We remark that our effective one-hole Lagrangian, Eq.~\re{l-one-hole}, 
differs significantly from the effective Lagrangian that has been derived by  
Shankar in Ref.~\onlinecite{S89,S90} for the same physical situation. 
Partially, the difference stems from the fact that the work in 
Ref.~\onlinecite{S89,S90} is based on a generalised $t$-$J$ model 
which includes besides nearest neighbour (intersublattice) hopping also 
next-nearest neighbour (intrasublattice) hopping.
Furthermore, however, terms that describe intersublattice hopping, i.e., 
the terms in the last line of Eq.~\re{l-one-hole}, 
are negelected in Ref.~\onlinecite{S89,S90}. 
It is argued that intersublattice hopping processes are necessarily  
accompanied by spin fluctuations and are therefore supressed in a situation 
with strong short-range antiferromagnetic order. 
In our case of a $t$-$J$ model with purely nearest neighbour hopping, 
the neglect of the corresponding terms in the effective Lagrangian 
would leave us with a static hole which is clearly not an appropriate
description of the physical situation which we intend to consider.

\section{Green's function}

In order to obtain the quasi--particle weight $Z$ of a hole in an 
antiferromagnetic background, we study the Green's function of a single hole 
in the $t$-$J$ model

\be
  G_{\sigma , \sigma '}(\vr - \vrs; \tau) \;=\; 
  \langle \; c_{\vr,\sigma}^{\dagger}(\tau) \; 
             c_{\vrs \!,\sigma '}^{}(0) \; \rangle
\ee

As is seen from the spectral representation, one can extract the 
quasi--particle weight from $G$ in the limit of large imaginary time $\tau$ 
at zero temperature:

{\arraycolsep-0.5ex
\bea
 &&       G_{\sigma , \sigma}(0; \tau \to \infty) \; \Big|_{T=0} \NL
  &&\sim \; \frac{1}{L^d} \; \sum_{\vq} \; 
  e^{- (E_{\vq}^{tJ} - E_0^{Heis.}) \, \tau} \;
  \left|\; {}_{tJ}\langle \vq |  c_{\vq\sigma} | 0 \rangle _{Heis.}\; \right|^2
 \;.  \label{spectral-rep}
\eea}

Here, $| \vq\rangle _{tJ}$ denotes the ground state of a single hole with 
momentum $\vq$ in the $t$-$J$ model, $| 0 \rangle _{Heis.}$ the ground
state of the Heisenberg model, and $E_{\vq}^{tJ}$ and $E_0^{Heis.}$ the 
corresponding energies; $L$ is the linear size of the system and $d$ its 
dimension. 
The last factor in \re{spectral-rep} represents the quasi--particle weight, 
and we are interested, in particular, in its size dependence. 

Our treatment of the Green's function is based on a path-integral 
representation for $G_{\sigma , \sigma '}(0; \tau)$. 
As in the previous Section, we generalise our considerations to arbitrary 
spin $S$ and use the representation \re{c-Darstellung} of the Fermion operators 
as a product of a Grassmann variable and an $SU(2)$--matrix. 
Noting that $G_{\sigma , \sigma '}$ is diagonal in the spin--indices, 
$G_{\sigma , \sigma '}(0; \tau) =: 
 \delta_{\sigma , \sigma '} \, S \, G(\tau)$,  
we arrive at the representation

{\arraycolsep0.3ex
\bea
G(\tau) = 
\frac{1}{{\cal Z}} \int \! {\cal D}[\omega] 
 &&  \eta_{\vr}^{}(\tau)  \eta_{\vr}^*(0)  \;
 \left[ g_{\vr}^{\dag}(\tau) g_{\vr}^{}(0)\right]_{\up \up} \NL
 && e^{-\int_0^{\beta}d\tau \;
 \left\{ \oml \dtau \omr \;+\; \oml{\cal H} \omr \right\} }  \;\; .
\eea}

$G$ is independent of $\vr$, so that we can choose $\vr=0$ which we consider as 
the origin of the A--sublattice. 
Following the steps that led to the effective action in Section II, 
we decompose the lattice into plaquettes and integrate 
over the $SU(2)$--field $M_{\vx}$ which is defined within each plaquette.
Next, we restrict the effective Lagrangian to zero--hole and one--hole terms, 
$\Lnls + \Loneh$, see above. 
Performing then the integration over the Grassmann variables $\eta_{\vr}^{}$ 
and $\eta_{\vr}^*$, we finally arrive at the formal result
\be
 G(\tau) = \frac{\int \! {\cal D}[g] \;
  \Goneh [g;\tau] \, 
 \left[ g_{\bm{0}}^{\dag}(\tau) g_{\bm{0}}^{}(0)\right]_{\up \up} \, 
   e^{-\int_0^{\beta}\!d\tau {\cal L}_{\mbox{\scriptsize nl$\sigma$m}} }}
 {\int \! {\cal D}[g] \; 
   e^{-\int_0^{\beta}\!d\tau {\cal L}_{\mbox{\scriptsize nl$\sigma$m}} }} \,. 
   \label{g}
\ee

Here, $\Goneh$ is the $\vr\!=\!\vrs\!=\!0$ element of the hole Green's function 
in matrix representation, calculated for a fixed configuration 
of the $SU(2)$--field $g$ that describes the spins, 
\be
 \Goneh [g;\tau] \;=\; \left[ 
  T \left\{ e^{- \int_0^{\tau} d\tau' \; \bm{h}[g]} \right\} \right]_{\bm{00}} 
\label{gg} \;.
\ee

This expression results as the solution of the corresponding equation 
of motion. 
$T$ is the time ordering symbol and 
$\bm{h}[g]$ is defined as the matrix (on the spatial lattice) which connects 
the Grassmann variables $\eta^*$, $\et^*$ and $\eta$, $\et$ in the quadratic 
form in $\Loneh$, cf.~Eq.~\re{l-one-hole}.
In the derivation of $G(\tau)$, Eq.~\re{g}, we have restricted not only 
the action but also the integration over the Grassmann variables 
to zero--hole and one--hole terms; 
in addition, we consider only the leading order in the limit $T\!\to\!0$. 
Effectively, this implies the omission of any correction proportional to the 
fugacity of the hole, $e^{-DJS/T}$; here, $DJS$ is the creation energy 
of one hole in a perfect antiferromagnetic background, cf.~Eq.~\re{h-diag}.

The general result for the Green's function, $G(\tau)$, 
Eqs.~(\ref{g}, \ref{gg}), for one hole in the $t$-$J$ model is valid 
under the conditions stated in connection with the derivation 
of the effective action, cf.~the text after Eq.~\re{h-ausser-diag}: 
in other words, it is valid whenever there is local antiferromagnetic order 
with a sufficiently large correlation length so that lattice effects 
play no role. 

Hole- and spin- degrees of freedom are coupled in $G(\tau)$, because the 
hole Green's function $\Goneh [g;\tau]$ depends on the time- and spatial 
fluctuations of the spin configuration, 
cf.~Eqs.~(\ref{h-diag}, \ref{h-ausser-diag}). 
A further evaluation of $G(\tau)$ in the presence of this coupling 
would require the exact calculation of $\Goneh [g;\tau]$ for each individual 
configuration of the $SU(2)$ background field $g_{\vx}(\tau)$. 
This is an impracticable task. 
Therefore, to be able to proceed we have to take recourse to approximations. 

We start from the expression Eq.~\re{g} for $G(\tau)$.
In the limit $\tau\! \to \!\infty$, $G(\tau)$ has the spectral representation 
(cf.~Eq.~\re{spectral-rep})    
\be
 G(\tau \to \infty) \; \sim \; 
  \frac{1}{{\cal N}} \; \sum_{\vq} \; 
  e^{- E_{\vq}(S) \, \tau} \; Z_{\vq}(S) \;, \label{spectr-res}
\ee
where $E_{\vq}(S)$ and $Z_{\vq}(S)$ are the hole energy and the 
quasi--particle weight of the hole, respectively. 

If we neglect the fluctuations of the spin configuration completely in 
Eq.~\re{g}, $g_{\vx}(\tau) = const.$, Eqs.~(\ref{gg}, \ref{h-diag}) yield 
$\Goneh [g=const.;\tau] = e^{-DJS\,\tau}$, 
and $E_{\vq}(S)=DJS$ and $Z_{\vq}(S)=1$ in this case. 
If we would neglect merely the coupling to the spin fluctuations in $\Goneh$,
we would obtain a non--trivial result for $Z_{\vq}(S)$.
In this case, the factorization of the hole- and the spin-part of $G(\tau)$ are 
reminiscent of the spinon--holon decomposition observed in one dimension.

The amplitudes of the spin fluctuations are controlled by the parameter $1/S$. 
In the sequel, we shall study the Green's function $G(\tau \to \infty)$,  
Eq.~(\ref{spectr-res}), in an expansion in powers of this parameter, i.e., 
in the semiclassical expansion:
$E_{\vq}(S)=DJS + {\cal O}(S^0)$, $Z_{\vq}(S)=1 + {\cal O}(S^{-1})$. 
The technical details of this expansion are presented in the 
Appendix \ref{SWEntwicklung}. 
We are interested in the quasi--particle weight only. 
Therefore, we disregard all corrections to the hole energy beyond 
the lowest order. 
This is achieved by scaling $\tau = {\tilde \tau} /S$ and neglecting all 
terms proportional to powers of $\tilde \tau$ which would result from 
corrections to the lowest order of $E_{\vq}(S)$ in Eq.~\re{spectr-res} 
in a systematic expansion in $1/S$. 
Then, our procedure yields an average of just the quasi--particle weight 
over the Brillouin zone:
\be
 Z(S) \;=\;  \frac{1}{{\cal N}}  \sum_{\vq}  Z_{\vq}(S)
 \;=\;   1 - \frac{1}{S}  Z^{(1)}
   - \frac{1}{S^2} Z^{(2)} + {\cal O}(\frac{1}{S^3}) . \label{res}
\ee

The fact that we cannot resolve the $\vq$--dependence of the quasi--particle 
weight within an $1/S$ expansion, is easy to understand:
as it is shown in the Appendix \ref{SWEntwicklung}, the expansion in 
$1/S$ also implies an expansion in the hopping amplitude $t$.
Without hopping, $h^{ab}=h^{ba}=0$ in Eq.~\re{l-one-hole}, $\Loneh$ 
becomes local in the hole fields, and thus, the hole energy $E_{\vq}(S)$ 
in Eq.~\re{spectr-res} becomes dispersionless in leading order 
in the $1/S$ expansion so that we arrive at the average. 

Our explicit results for the first two coefficients $Z^{(1,2)}$ 
of the $1/S$ expansion of the quasi--particle weight in one and two i
dimensions will be presented next.

\section{Results and Summary}
 
We start with the discussion of the one--dimensional case. 
The $k$--sums in the results for $Z^{(1,2)}$, 
Eqs.~(\ref{res1}, \ref{res2}) in the Appendix \ref{SWEntwicklung}, 
are performed as integrals over the Brillouin zone with a 
cutoff at small momenta, $k \gtrsim k_0=2\pi/L$.
Then, $Z^{(1)}$ diverges logarithmically with the system size $L$:

\be
 Z^{(1)} \;=\; (1+\frac{4 t^2}{J^2}) \; \frac{1}{2\pi} \ln{\frac{L}{a}} \;.
\ee

Here, we have neglected terms which remain finite in the limit 
$L \to \infty$ since they depend on the choice of the cutoff. 
For $S=1/2$, the quasi--particle weight $Z$ of a single hole is known 
to vanish algebraically with increasing system size $L$, $Z \sim L^{- 2X}$ 
(see Refs.~\onlinecite{SP96}, \onlinecite{SP98}).
This behavior of $Z$ reflects Anderson's orthogonality catastrophe 
\cite{A67a,A67b}:
the states $c_{\vq\sigma} | 0 \rangle _{Heis.}$ and $ | \vq \rangle _{tJ}$ 
(see Eq.~\re{spectral-rep}), i.e., the bare one-hole state and 
the true one-hole eigenstate of the $t$-$J$ model, are orthogonal 
in the thermodynamic limit. 
The orthogonality catastrophe is an effect of the quantum fluctuations, 
and therefore, it should not occur in the classical limit $S\to\infty$. 
Consequently, we expect that $\lim_{S \to \infty} X = 0$. 
This suggests that the expansion in Eq.~\re{res} should be 
reformulated as an expansion of $\ln Z(S)$ in powers of $1/S$. 
Then, we get in first order for the exponent $X$ the result 
$X=(1+4 t^2/J^2)/(4\pi S)$. 

At first, this looks quite satisfactory:
the exact value for $S=1/2$ at $t=0$ is $X=3/16$ (see Ref.~\onlinecite{SP98}). 
As $t$ increases from zero, $X$ also increases. 
This tendency is compatible with another exactly known result: 
at the supersymmetric point $J/t=2$, where the $t$-$J$ model is 
exactly solvable, one finds value $X^{Susy} = 1/4$; 
see Ref.~\onlinecite{SP96}. 

However when we try to improve the lowest order result by including 
the next order

\be
 Z^{(2)} \;=\; 
  (\frac{1}{4} - \frac{4 t^2}{J^2})
  \left( \frac{1}{2\pi} \ln{\frac{L}{ a}} \right)^2   
 + {\cal O}\left( \ln{\frac{L}{ a}} \right) \;,
\ee

we find in leading logarithmic order 

\be
 \ln Z(S) \;=\; -(1+\frac{4 t^2}{J^2}) \; \frac{1}{2\pi S} \ln{\frac{L}{4 a}}\;
  - \frac{3}{4} \left( \frac{1}{2\pi S} \ln{\frac{L}{4 a}} \right)^2  \;.
\label{lnZ}  
\ee

Here we have omitted a term of the order of $\frac{1}{S^2} (t/J)^4$, since 
it does not change the picture qualitatively.
Obviously, this expression for $\ln Z(S)$ cannot simply be interpreted 
as the first two terms of an expansion in powers of $1/S$ of the exponent 
$X(S)$ in a power law $Z(S) \sim L^{-2X(S)}$. 
At first glance, one might think that the deviation from such an expansion 
can be attributed to the fact that in Eq.~\re{res}, the quantity $Z(S)$ 
is an average over the Brillouin zone of the quasi--particle weights 
$Z_{q}(S)$ so that one cannot expect to find a power--law dependence 
of $Z(S)$ on the system size $L$. 
However, for $t\!=\!0$, i.e.~for a static hole, for which $Z_{q}(S)$ 
is independent of the wave number $q$, Eq.~(\ref{lnZ}) can still not 
be interpreted as an expansion in powers of $1/S$. 
The infrared divergent terms in $Z(S)$ show that instead of a straightforward 
expansion a renormalization group treatment is necessary. 
This is quite plausible, since $1/S$ is the coupling constant of the 
non--linear $\sigma$--model $\Lnls$, Eq.~(\ref{nlsm}), 
on which our $1/S$ expansion is based. 
However, in Eq.~\re{g} the parameter $1/S$ appears not only as the coupling 
constant of $\sigma$--model $\Lnls$, it also occurs in the hole Green's 
function, Eq.~\re{gg}. 
Therefore, the standard renormalisation group procedure for the non--linear 
$\sigma$--model \cite{Amit84} cannot be applied in the present case.
The development of an appropriate procedure is beyond the scope of this paper. 
Notwithstanding these complications, Eq.~\re{lnZ} shows a behavior compatible 
with a vanishing quasi--particle weight in one dimension. 
  
In two dimensions, our general expressions  Eqs.~(\ref{res1}, \ref{res2}) 
for $Z^{(1,2)}$ yield a different picture: 
the $\vk$--sums converge at small momentum $k$. 
Thus, in contrast to the one--dimensional case, the average quasi--particle 
weight $Z(S)$ is independent of the system size. 
This means that up to the order $1/S^2$ of our large-$S$ expansion 
there is no sign of an orthogonality catastrophe in the two-dimensional model. 
In recent, very elaborate numerical studies of  $t$-$J$ model on the square 
lattice \cite{BAM00b,MPS01} it has been found that within this model, 
the spectral function of a single hole shows the signature of a coherent 
quasi--particle, i.e., the quasi--particle weight $Z_{\vq}(S)$ has been 
found to remain finite throughout the Brillouin zone.  
Although the semiclassical expansion does not allow us to determine 
the weights $Z_{\vq}(S)$ at individual $\vq$ points, 
our result for the average weight $Z(S)$ is consistent with these 
numerical findings.

In summary, in this paper we have presented a new approach to the problem 
of hole dynamics in the $t$-$J$ model. 
We use a generalisation of this model, which has originally been 
designed for spin $\frac{1}{2}$ particles, to describe particles 
with arbitrary spin $S$. 
This has allowed us to consider the case of large $S$ in which the 
semiclassical approximation is applicable. 
This approximation has the advantage that it works independent of the 
spatial dimension of the system we wish to investigate. 
Therefore, we have been able to study the single--hole dynamics 
in one and in two dimensions on the same footing. 
In agreement with most previous investigations of the dynamical properties 
of a hole in the $t$-$J$ model, we find a qualitative difference 
between these properties in one and in two dimensions: 
while in two dimensions the results of our $1/S$ expansion are compatible 
with the picture of the hole as a coherent quasi-particle in the sense 
of Landau's Fermi liquid theory, 
the hole appears to have a vanishing quasi--particle weight in the one 
dimensional case so that the quasi--particle picture fails in this case.

\appendix

\section{Coherent States}
\label{Koh-Zust}
We use the method of coherent states (for a recent review see 
Ref.~\onlinecite{ZFG90}) in order to deal with the problem of 
excluded double occupancy \cite{W88}.
First, consider the case of spin $1/2$. 
Then, the Hilbert space consists of a fermionic state with spin up 
$c_{\up}^{\dag}| 0 \rangle$, a fermionic state with spin down 
$c_{\dwn}^{\dag}| 0 \rangle$, and a hole $| 0 \rangle$ 
at each lattice site. 
Following Ref.~\onlinecite{AEK98}, we introduce (at each lattice site) 
the states

\bea
\omr  &=&  
  e^{\textstyle \left[ \eta \; c - c^{\dag} \; \eta^*  \right]}  \;
 c^{\dag} \; | 0 \rangle \NL
 &=& (1 - \halb \eta^* \eta^{}) \; c^{\dag} \; | 0 \rangle
 \;+\; \eta \; | 0 \rangle         \label{omega}
\eea
with
\be
 c^{\dag} \;=\; c_{\up}^{\dag} \; g_{\up \up} \;+\; 
	       c_{\dwn}^{\dag} \; g_{\dwn \up}   \;.
\ee

Here, the parameters $\eta$, $\eta^*$ are Grassmann variables and 
describe the hole.
$g$ is a $SU(2)$--matrix, $g^{\dag} \, g = \eins{2}$; it can be parametrized 
by three Euler angles and describes the orientation of the spin.
In this Appendix, we index a unit matrix with its dimension. 

Next, we generalize the coherent states $\omr$ to the case of arbitrary 
spin $S$.
Then, the Hilbert space consists of $(2S\!+\!1) \,+\, 2S$ states 
at each lattice site which are the states of a spin 
$S$, $|S, m=-S \cdots S \rangle$ and a spin 
$\Sm$, $|(\Sm), m=-(\Sm) \cdots (\Sm)\rangle$.
Now define, cf.~Ref.~\onlinecite{AL91}, 
\be
\omr \;=\; (1 - \halb \eta^* \eta^{}) \; R \; |S,S\rangle 
              \;+\; \eta \; R \; |(\Sm),(\Sm)\rangle  \label{omegaS} \;.
\ee
Here, the rotation matrix $R$ can be represented for arbitrary spin 
in the following way by three Euler angles $\psi$, $\theta$, and $\phi$
\be
 R \;=\; e^{-i\phi\hat{S}^z} \; e^{-i\theta\hat{S}^y} \; 
         e^{-i\psi\hat{S}^z} \;.
\ee
Instead of using the Euler angles, one can also parametrize $R$ by the 
elements of the $SU(2)$--matrix $g$ 
which is the $S=\halb$ representation of $R$. 
In this paper, we use $g$ as the fundamental variable. 
In the case $S=\halb$, \re{omegaS} coincides with \re{omega}. 

For the application of the method of coherent states, we need to specify 
the expectation values of the operators in the Hamiltonian 
with the states $\omr$.
Obviously, $\oml \omr =1$.
The spin operator $\hat{\bf S}$ acts as usual on the spin--states; 
with $s=S$ or $s=\Sm$:
\be
 \langle s,s| \; R^{\dag} \; \hat{\bf S} \; R \; |s,s\rangle 
 \;=\; s \; \vn_{}  \;.
\ee
In terms of $g$, the unit vector $\vn$ reads
\be
 \vn_{} \;=\; \Sp{ \bm{\sigma} \; g_{}^{} \sigma_z g_{}^{\dag} } \;.
\ee
For the expectation value of the spin with $\omr$, we find
\be
 \oml \hat{\bf S} \omr \;=\; \halb \; \ho{} \; \vn_{} \;.
\ee
The Fermion operators $c_{\sigma}^{}$ 
(we use $\sigma =\; \up,\dwn\; \equiv \; \pm 1$) link the states with spin 
$S$ and $\Sm$, 
\be
 c_{\sigma} \; |S, m\rangle \;=\; \gamma_{\sigma , m} \; 
 |\Sm, m-\sigma \halb \rangle  \;. 
\ee
Here, $\gamma_{1, -S} = \gamma_{-1, S} =0$ and the remaining 
coefficients $\gamma_{\sigma , m}$ are uniquely determined by demanding
that the relation between the Fermion operators and the spin is as 
in the case of $S=\halb$, 
\be 
 \halb \left(c_{\up}^{\dag} c_{\up}^{} - c_{\dwn}^{\dag} c_{\dwn}^{} \right) 
  \;=\; \hat{S}^z  \;\;\mbox{and} \;\;  
  c_{\up}^{\dag} c_{\dwn}^{} \;=\; \hat{S}^+  \;.
\ee
We get 
\be
 \oml c_{\sigma}^{} \omr \;=\; \sqrt{2S} \; \eta^* \; g_{\sigma \up} \;.
 \label{c-Darstellung}
\ee
The resolution of unity reads in our case
\be
 \int d\omega \; \omr \oml \;=\; \eins{4S\!+\!1} \;.
\ee
The integration measure is
\be
\int d\omega \; \dots \;=\; \int dg\int d\eta^* d\eta^{} \; 2S \;
     e^{-\frac{1}{2S} \eta^* \eta^{} } \; \dots  \;\;.   \label{domega}
\ee
Our convention $\int dg =1$ for the invariant measure of $g$ leads to 
\be
 \int dg \;  R \; |s,s\rangle \langle s,s| \; R^{\dag} \;=\; 
   \frac{1}{2s+1} \;  \eins{2s\!+\!1}  \;\;.  \label{degeh}
\ee
With the aid of Eqs.\ (\ref{omegaS},\ref{domega},\ref{degeh}), one 
quickly confirms that the dimension of the Hilbert space comes out correctly:
{\arraycolsep-1ex
\bea
 && {\rm Tr}\left\{ \int d\omega \omr \oml \right\}     \NL
 &&= 
 \int d\eta^* d\eta^{} \; 2S \; e^{-\frac{1}{2S} \eta^* \eta^{} } \;
 (1-2\eta^* \eta^{}) \;=\; 4S\!+\!1 \;.
\eea}
To complete the list of relations involving the coherent states, 
we quote the expression entering the kinetic part of the action 
(we consider now $\omega$ to depend on $\tau$)
\be
 \oml \dtau \omr \;=\;  \ho{} 
 \left( g_{}^{\dag} \dtau g_{}^{} \right)_{\up \up}
 \;+\; \eta_{}^* \dtau \eta_{}^{} 
 \;-\; \halb \dtau (\eta_{}^* \eta_{}^{})  \label{kinetic}
\ee
In deriving Eq.\ \re{kinetic}, we used
\be
 \langle s,s| \; R^{\dag} \; \dtau \; R \; |s,s\rangle
 \;=\; 2s \; \left( g_{}^{\dag} \dtau g_{}^{} \right)_{\up \up} \; \;.
\ee

\section{$1/S$ expansion}
\label{SWEntwicklung}
Here, we describe the simultaneous expansion of $G(\tau)$, 
Eq.~\re{g}, in the fluctuation amplitudes and in the hopping integral $t$.
The antiferromagnetically ordered state, $\vn_{\vx}(\tau)= \hat{e}_z$ 
is given by $g_{\vx}(\tau)=1$ (cf.~Eq.~\re{nvec}). 
Then, it is convenient to characterize an arbitrary state by
\be
 g_{\vx}(\tau) \;=\; e^{i \bm{\sigma}\eps(\vx, \tau)} \;
                      e^{i \sigma_z \epsilon_z(\vx, \tau)} \; , \label{SWpara}
\ee
where we parameterize the field $g_{\vx}(\tau)$ by a $SU(2)/U(1)$--symmetric 
part and a $U(1)$--factor.

It is easily seen that the latter disappears from our considerations:   
$\vn_{\vx}(\tau)$ is determined only by the two fields in the vector 
$\eps = (\epsilon_x,\epsilon_y)$ which enter the $SU(2)/U(1)$--part 
of the field $g$; the $U(1)$--part, the last factor in \re{SWpara} drops out. 
Thus, the $U(1)$--factor also drops out of $\Lnls$. It still appears in 
$\Loneh$, Eq.~\re{l-one-hole}, through the field 
\be
A^{\tau}_{\vx \, \sigma \sigma}(\tau) \;=\; \sigma\,\dtau\,\epsilon_z(\vx, \tau)
  -i \left[e^{-i \bm{\sigma}\eps(\vx, \tau)} \;\dtau\; 
	 e^{i \bm{\sigma}\eps(\vx, \tau)} \right]_{\sigma \sigma}
\ee
and through the off-diagonal matrix elements $h^{ab}_{\vx+\vd,\vx}$ and 
$h^{ba}_{\vx,\vx+\vd}$ (cf.~Eq.~\re{h-ausser-diag}) in which 
the $U(1)$--terms occur as the first and the last factor. 
Looking at Eq.~\re{l-one-hole} one recognises, however, that these  
$U(1)$ factors can be absorbed into the Grassmann fields by the gauge 
transformation 
\be
\eta_{\vx}(\tau) \to  e^{i \epsilon_z(\vx, \tau)} \; \eta_{\vx}(\tau)
\;, \;\;\;
\et_{\vx}(\tau) \to  e^{-i \epsilon_z(\vx, \tau)} \; \et_{\vx}(\tau)\;,
\ee

so that they disappear from the integrands in the numerator and 
in the denominator on the r.~h.~s.~of Eq.~\re{g}. 
Furthermore, in the integration measure $\epsilon_z$ and $\eps$ separate, 
$dg \propto d \epsilon_z \; d^2 \epsilon \sin(2\epsilon)/\epsilon$, 
so that the integrations over $\epsilon_z$ cancel between the numerator and 
the denominator of the r.~h.~s.~ of  Eq.~\re{g}.
These considerations show that the field $\epsilon_z$ disappears completely 
from the expression  Eq.~\re{g} for the Green's function $G(\tau)$.

Now, we consider this expression in a systematic expansion with respect to 
$\eps$, i.e. we expand on  the r.~h.~s.~of Eq.~\re{g} the integration measure, 
the factors $\Goneh$ and $[g_{\bm{0}}^{\dag} g_{\bm{0}}^{}]_{\up \up}$ 
in the integrand, and $\Lnls$ in powers of $\eps$. 
Inspection of the quadratic part of $\Lnls$ shows that in this expansion, 
each additional power of $\eps$ beyond the second order yields 
an additional power of $1/\sqrt{S}$. 
We make this explicit by scaling the spin wave amplitude $\eps$ by a factor 
of $(4S)^{-1/2}$, $\eps = \epst /(4S)^{1/2}$ and the imaginary time 
by a factor of $(2dJS)^{-1}$, $\tau = \tilde{\tau}/(2dJS)$. 
Then, the quadratic part in the new fields becomes independent of $S$ and 
the spin wave dispersion reads 
\be
 \oq \;=\; \sqrt{1 - \gq ^2} \;, \mbox{with}\;\;
    \gq \;=\; \frac{1}{d}\sum_{s=1}^d \cos(k_s a)   \;.
\ee
Relative to the leading order in $S$, $G(\tau) = \exp(-\tilde{\tau}/2)$, 
we want to obtain corrections up to and including the order $1/S^2$ 
accurately. 
Therefore, we retain terms up to the order $\epst ^4$ in the factors 
$\Goneh$ and $[g_{\bm{0}}^{\dag} g_{\bm{0}}^{}]_{\up \up}$ 
in the integrand, anharmonicities up to the order $\epst ^4$ in $\Lnls$, 
and terms up to the order $\epst ^2$ in the measure 
(note that after the scaling, the measure already contains a prefactor 
of $1/S$ in the order $\epst ^2$). 
In performing the integrals over the various combinations of powers 
of $\epst$, which occur after this expansion in the numerator and 
the denominator of Eq.~\re{g}, use can be made of the linked-cluster theorem.

$\Goneh$ contains spatially diagonal, $h^{aa}$ and $h^{bb}$, as well as 
spatially off--diagonal terms, $h^{ab}$ and $h^{ba}$.
The latter are of the order of $\eps \sim 1/S^{1/2}$.
Since they are proportional to the hopping integral $t$ 
(cf.~Eq.~\re{h-ausser-diag}), our expansion in $1/S$ becomes necessarily 
also an expansion in $t$.
We shall neglect terms of order higher than $t^2$. 

The actual expansion is straight--forward but rather tedious. 
We express the result as
\be
 G(\tau) \;=\; e^{-\tilde{\tau}/2} \,
 \left( 1 - \frac{1}{S}  G^{(1)}(\tilde{\tau})
   - \frac{1}{S^2} G^{(2)}(\tilde{\tau}) + {\cal O}(\frac{1}{S^3})\right) \,.
\ee

For the first order coefficient, we obtain
\be
 G^{(1)}(\tilde{\tau}) \;=\; - \cdots \, \tilde{\tau} 
  + \sumk{{}}{\vk} \frac{1- e^{-\oq \tilde{\tau}}}{4 \oq } \; 
   (1+\frac{4 t^2}{dJ^2})  \;.          \label{G-eins}
\ee
The first term in Eq.~\re{G-eins}, proportional to $\tilde{\tau}$, 
belongs to the expansion of $E_{\vq}(S)$ in Eq.~\re{spectr-res} 
and will consequently be neglected. 
In the evaluation of the momentum sum in the second term in Eq.~\re{G-eins}, 
we consider a finite lattice of $\cal N$ lattice points of the A--sublattice 
and exclude the point $\vk=0$ from the sum, since the corresponding mode 
is a Goldstone mode, the uniform rotation of all spins.  
Then, this term has a finite limit as $\tau \to \infty$ and defines $Z^{(1)}$,
the first order correction of the quasi--particle weight in Eq.~\re{res} as 
\be
 Z^{(1)} \;=\; \sumk{{}}{\vk \ne 0} \frac{1}{4 \oq} \; 
                (1+\frac{4 t^2}{dJ^2}) \;.  \label{res1}
\ee 

The calculation of the next order coefficient, $G^{(2)}(\tilde{\tau})$, 
proceeds exactly in the same way, but it is much more tedious. 
We neglect polynomial terms $\tilde{\tau}$, $\tilde{\tau}^2$, since they 
belong to the expansion of $E_{\vq}(S)$ in Eq.~\re{spectr-res}, 
exclude the point $\vk=0$ from the momentum sums, and take the limit 
$\tau \to \infty$.
Then finally, collecting all terms, we get the following result for the second 
order correction in the quasi--particle weight 

\be
 Z^{(2)} \;=\; 
 \sumk{2}{{\vk \ne 0}\atop {\vk' \ne 0}} \frac{1}{4 \oq} \frac{1}{4 \oqs}  
 \left( \! A_{\vk , \vk'} + \frac{4 t^2}{dJ^2} \; B_{\vk , \vk'} \! \right) 
 \; , \label{res2}
\ee 

with
\bea
 A_{\vk , \vk'} &=& \frac{1}{4} 
 - \frac{\oq \oqs}{\oq + \oqs} \;, \\*
 B_{\vk , \vk'} &=&  -1 + \frac{\oq \oqs}{(\oq + \oqs)^2}  \NL
&& + (\oq + \oqs )
  \left( \frac{3}{4} + \left[ \frac{\gqp - \gq \gqs}{\oq \oqs } \right]^2 
   \right)  \NL
&& - \frac{1}{2} \, \frac{\oq \oqs}{\oq + \oqs}  
  \left( 3 + \left[ \frac{\gqp - \gq \gqs}{\oq \oqs } \right]^2  \right)  \;.  
\eea

In the preceeding expressions for $A_{\vk , \vk'}$ and $B_{\vk , \vk'}$, 
we have neglected terms of order $k^2$, $k'^2$, or $\vk \vk'$ and higher.

\end{document}